\documentclass[sigconf, nonacm]{acmart}

\usepackage{subcaption}

\usepackage{tcolorbox}
\usepackage{MnSymbol}
\usepackage{tikz}

\AtBeginDocument{%
  }

\begin{document}

\title{Insights from the Usage of the Ansible Lightspeed Code Completion Service}
\titlenote{This paper has been published at the 39th IEEE/ACM International Conference on Automated Software Engineering (ASE 2024), Industry Showcase under the title "Ansible Lightspeed: A Code Generation Service for IT Automation".}

\author{Priyam Sahoo}
\affiliation{%
  \institution{Red Hat}
  \city{}
  \country{}
  }

\author{Saurabh Pujar}
\affiliation{%
  \institution{IBM Research}
    \city{}
  \country{}
  }
\email{saurabh.pujar@ibm.com}

\author{Ganesh Nalawade}
\affiliation{%
  \institution{Red Hat}
    \city{}
  \country{}
  }
\email{gnalawad@redhat.com}

\author{Richard Gebhardt}
\affiliation{%
  \institution{Red Hat}
    \city{}
  \country{}
  }

\author{Louis Mandel}
\affiliation{%
  \institution{IBM Research}
    \city{}
  \country{}
  }

\author{Luca Buratti}
\affiliation{%
  \institution{IBM Research}
    \city{}
  \country{}
  }

\begin{abstract}

The availability of Large Language Models (LLMs) which can generate code, has made it possible to create tools that improve developer productivity.
Integrated development environments or IDEs which developers use to write software are often used as an interface to interact with LLMs.
Although many such tools have been released, almost all of them focus on general-purpose programming languages.
Domain-specific languages, such as those crucial for Information Technology (IT) automation, have not received much attention.
Ansible is one such YAML-based IT automation-specific language.
Ansible Lightspeed is an LLM-based service designed explicitly to generate Ansible YAML, given natural language prompt.

In this paper, we present the design and implementation of the Ansible Lightspeed service. 
We then evaluate its utility to developers using diverse indicators, including extended utilization, analysis of user edited suggestions, as well as user sentiments analysis.
The evaluation is based on data collected for 10,696 real users including 3,910 returning users. 
The code for Ansible Lightspeed service and the analysis framework is made available for others to use.

To our knowledge, our study is the first to involve thousands of users of code assistants for domain-specific languages.
We are also the first code completion tool to present N-Day user retention figures, which is 13.66\% on Day 30.
We propose an improved version of user acceptance rate, called Strong Acceptance rate, where a suggestion is considered accepted only if less than $50\%$ of it is edited and these edits do not change critical parts of the suggestion.
By focusing on Ansible, Lightspeed is able to achieve a strong acceptance rate of 49.08\% for multi-line Ansible task suggestions.
With our findings we provide insights into the effectiveness of small, dedicated models in a domain-specific context.
We hope this work serves as a reference for software engineering and machine learning researchers exploring code completion.

\end{abstract}


\maketitle

% problems
\section{Introduction}
\label{sec: intro}

\begin{figure}[t]
    \centering
    \includegraphics[width=\linewidth]{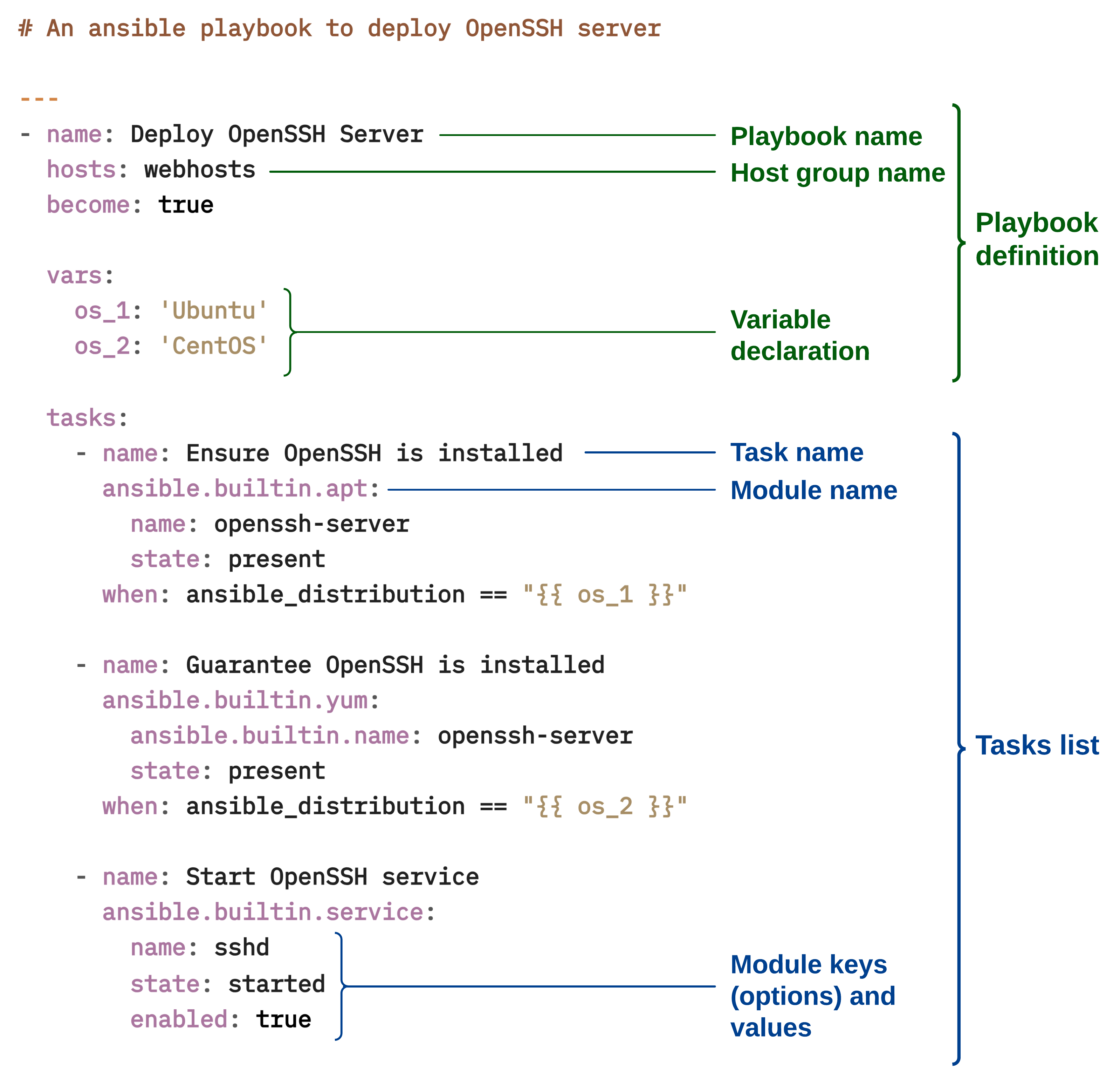}
    \vspace{-2em}
    \caption{A typical Ansible playbook structure consists of \emph{playbook definition} and \emph{tasks list}. Ansible Lightspeed generates a \emph{task}, given the \emph{task name} and the preceding context. \emph{Module name} is the first term generated by Lightspeed, followed by the module associated \emph{keys} and \emph{values}.}
    \label{fig:playbook_structure}
    \Description{}
\end{figure}

Ansible~\cite{ansible} is a YAML based domain-specific language dedicated to IT automation.
It is one of the most widely used infrastructure as code tools~\cite{guerriero19adoption,stackoverflow23adoption}.
Since it is open-source, we do not know the precise number of developers who use Ansible but we can get an indication of its popularity with the Ansible python community package~\cite{ansible_pypi} getting over 5 million downloads per month~\cite{ansible_pypi_stat}. 
Ansible also has an active community on GitHub~\cite{ansiblegithub} with about 5,000 contributors and about 54,000 commits on an yearly basis, which is likely a small fraction of the actual user base.\footnote{Another indication of Ansible's popularity is from Ansible's social media accounts of Reddit \cite{ansiblereddit} and X \cite{ansiblex}, counting 63,000 readers and 72,000 followers, respectively.}

A typical Ansible project is organized into \emph{playbooks}~(programs) and \emph{roles}~(libraries).
Figure~\ref{fig:playbook_structure} shows an Ansible playbook with it's sub-components.
Playbooks consist of \emph{plays}, which are a mapping between hosts and the tasks (sequential execution units) that run on those hosts.
Tasks contain a natural language description in the form of a \emph{name field}, a \emph{module name} defining the action to execute, and \emph{keys}~(or options) configuring the action. 

Code generation models, and some of the systems that utilize them, have emerged as powerful tools for software developers and system analysts~\cite{GitHubCopilot,Tab9,replit,codewhisperer}.
Studies show that AI tools for general-purpose programming languages improve productivity and also suggest that novice programmers may benefit more from such tools~\cite{weber24productivity,peng2023impact,barke2023oopsla}.
However, generic models do not perform as well on domain-specific languages.  \citet{pujar2023automated} showed that a relatively small model fine-tuned on high quality Ansible data can outperform a much larger and more general model on benchmark Natural Language to Ansible-task generation test data.
But this work does not involve real users in their development environment.

\begin{figure}
\centering
    \begin{subfigure}{0.5\textwidth}
        \includegraphics[width=\textwidth]{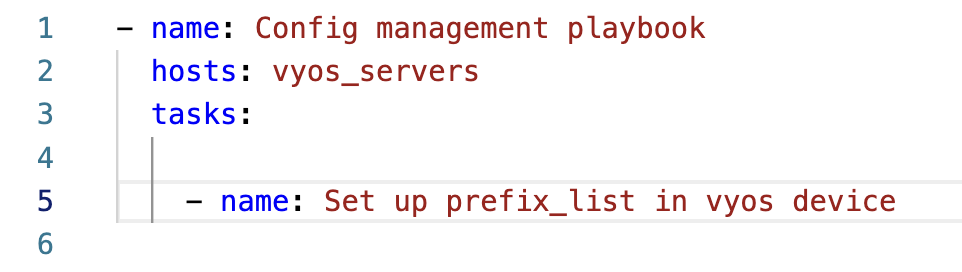}
        \caption{User writes the prompt.}
        \label{fig:ansible_lightspeed_flow_first}
    \end{subfigure}
\vspace{1em}
\vfill
\begin{subfigure}{0.5\textwidth}
    \includegraphics[width=\textwidth]{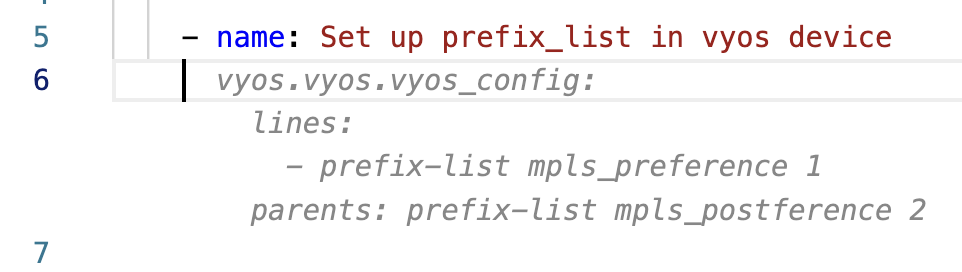}
    \caption{Ansible Lightspeed provides an inline suggestion.}
    \label{fig:ansible_lightspeed_flow_second}
\end{subfigure}
\vspace{1em}
\vfill
\begin{subfigure}{0.5\textwidth}
    \includegraphics[width=\textwidth]{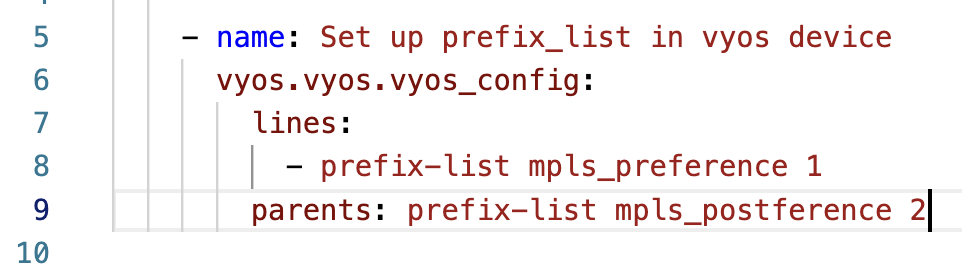}
    \caption{User accepts the suggestion by hitting the `Tab' key. }
    \Description{}
    \label{fig:ansible_lightspeed_flow_third}
\end{subfigure}
\caption{Ansible Lightspeed's workflow in the text editor. Users receive Lightspeed suggestions, which are almost always multi-line, after entering the task name and moving to the next line. Then, users can either accept the suggestion by pressing the ‘Tab’ key or reject it by pressing the ‘Esc’ key.}
\label{fig:ansible_lightspeed_flow}
\end{figure}

Red Hat Ansible Lightspeed~\cite{lightspeed}, further referred to as Ansible Lightspeed or simply Lightspeed, is a generative AI service that utilizes watsonx Code Assistant for Ansible (WCA), an extension of Ansible Wisdom~\cite{pujar2023automated}, to produce code recommendations based on best practices.
An example of a user interaction with Ansible Lightspeed is presented in Figure~\ref{fig:ansible_lightspeed_flow}.
With Ansible Lightspeed, you can build an Ansible playbook step-by-step, by providing a natural language description of an Ansible task  using its name field. 
The description becomes a prompt for the model to generate the code of the task, which usually consists of the module name followed by module keys and values.

Ansible Lightspeed was available since April 15 2023 to a limited number of users as part of a closed beta release, and then generally to everyone as part of a free Tech Preview version from June 2023 ~\cite{lightspeed} until May 15 2024.
During this period thousands of developers interacted with the system.
In this paper we analyze the interactions and feedback of 10,696 users of the closed beta and free Tech Preview versions of Ansible Lightspeed, who consented to share their data.
To our knowledge, there has been no published study, with several thousand users, on the use of code assistants for domain-specific languages.

As a part of our study we analyze usage trends on a temporal basis, to understand when Ansible Lightspeed is used.
We see if users continue to use Ansible Lightspeed and if so, how regularly.
Most importantly, we check to what degree users accept suggestions made by Ansible Lightspeed.
This acceptance rate is a crucial comparison metric with other code completion tools in the industry.
It is also essential for us to understand how accurate the accepted suggestions were from the user's perspective.
This is done through an edit analysis of the accepted suggestions.
The edit analysis tells us how the user edited the model suggestion before using it.
This gives valuable insights on whether the model suggestion was truly useful and clues on how we can improve it.
As a result of edit analysis, we also get a more effective and stringent criteria for acceptance, where we consider a suggestion as accepted only if the user does not edit most or critical parts of the suggestion after accepting it.
Finally, we analyze and share user qualitative feedback of the Ansible Lightspeed service.

The paper makes the following contributions:
\begin{itemize}
\item A detailed description of the Ansible Lightspeed system and analysis framework. 
The code for Ansible lightspeed\footnote{https://github.com/ansible/ansible-ai-connect-service} and the analysis framework\footnote{https://github.com/ansible-community/lightspeed-analysis-framework} has been made available on GitHub.
\item We are the first code completion tool to share user retention statistics. This can be used as a baseline for similar tools in the future. 
\item We propose an improved acceptance criteria, called Strong Acceptance rate, that checks if the accepted suggestion was actually used by the user.
\item We show that Ansible Lightspeed's acceptance rate is higher than that of other, more general, code completion tools.
\end{itemize}

Given that we have a domain-specific model, targeting a well defined use case, we are able to achieve high user acceptance rates with a relatively small model of size 350 million parameters.
Our analysis of user made edits to model suggestions, presents insights on how users use the suggestion and how we can improve it.
This study also provides a reference point for software engineering and machine learning researchers investigating the adoption of code completion services, especially for domain-specific languages.

\section{Ansible Lightspeed}
\label{sec: lightspeed}

\definecolor{uicolor}{RGB}{191,191,191}
\definecolor{inferencecolor}{RGB}{210,209,235}
\definecolor{matchingcolor}{RGB}{167,232,235}
\definecolor{telemetrycolor}{RGB}{191,240,191}
\definecolor{servicescolor}{RGB}{255,255,255}
\begin{figure*}[t]
    % \centering
    \hspace*{-1.5em}
    \includegraphics[width=1\linewidth]{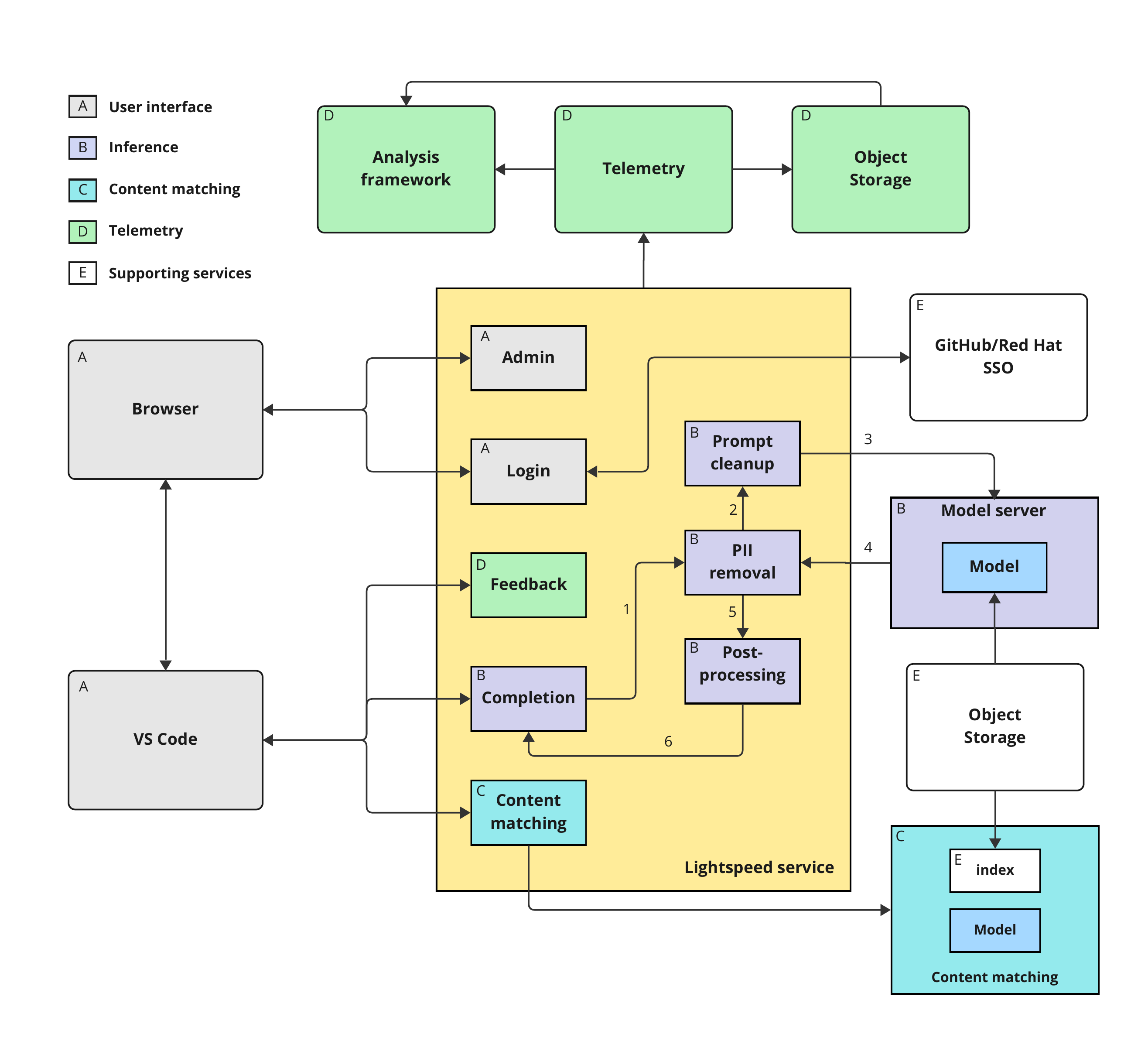}
    \caption{Ansible Lightspeed application architecture.
    The components 
    are:
    \fcolorbox{black}{uicolor}{\textnormal{\textsf{\footnotesize A}}} the user interface,
    \fcolorbox{black}{inferencecolor}{\textnormal{\textsf{\footnotesize B}}} the inference pipeline,
    \fcolorbox{black}{matchingcolor}{\textnormal{\textsf{\footnotesize C}}} the content matching pipeline,
    \fcolorbox{black}{telemetrycolor}{\textnormal{\textsf{\footnotesize D}}} the analysis framework and 
    \fcolorbox{black}{servicescolor}{\textnormal{\textsf{\footnotesize E}}} supporting services.
    The arrows indicate the information flow and the numbers indicate the processing order.
    \Description{}
    \label{fig:content_development_using_lightspeed}
    }
\end{figure*}

At a very high level, the Ansible Lightspeed platform consists of the following components:
\begin{itemize}
    \item A code editor extension that captures and transmits prompt and context information for inference and captures feedback from the user to improve model and service quality.
    \item An inference pipeline for combining and processing natural language and Ansible-YAML to get code suggestions from a large language model.
    \item A content matching pipeline for attribution, that finds training examples that are similar to the code suggestions and provides the location of the corresponding repositories.
    
    \item An analysis framework that collects and processes feedback data to make it consumable by a wide range of analysis tools, ultimately for the purpose of improving model quality and user experience.
\end{itemize}

Ansible Lightspeed is based on a client-server architecture, presented in Figure~\ref{fig:content_development_using_lightspeed}.
The client is the VS Code Ansible extension which is published on the Visual Studio Marketplace~\cite{ansibleextension} and provides a bulk of language specific features for Ansible.
The server consists of an application service that handles authentication, authorization and entitlement checking, exposes APIs for inference, feedback and content matching, processing of inference prompts and model recommendations to improve quality. 
It is designed to be highly scalable and highly available, deployed in a multi-zone and multi-region configuration, ensuring that a large number of requests can be served simultaneously and with minimal latency.

\subsection{Code Editor Extension}

The Ansible extension relies on the Ansible Language server~\cite{ansiblelanguageserver} that supports the Language server protocol~\cite{languageserverprotocol} to provide language specific features like code auto-completion, diagnostics, hover and go-to features.
The language service invokes Ansible Lint~\cite{ansiblelint} in the background to identify any issues with the Ansible file that the user is currently editing and provides real-time diagnostic information in the VS Code problems tab.

To use Ansible Lightspeed from within the code editor, the user must enable the Ansible Lightspeed feature in the Ansible extension settings, ensure the file type of the current file that's being edited is "Ansible" and then login to the Ansible Lightspeed service. After the login is successful, the user is ready to initiate requests and receive code suggestions from the Ansible Lightspeed service via the Ansible extension.

As shown in Figure~\ref{fig:ansible_lightspeed_flow}, when a user starts writing Ansible code with a task description of the form \texttt{-} \texttt{name:} \textit{<Natural language prompt>}, and then presses enter, a \textit{completion} request is sent to the Ansible Lightspeed service. The request includes all the content of the file up to the cursor position. After the server responds, the editor displays the response as greyed out text, which the user can either accept by pressing the tab key or reject by pressing the escape key.

For the accepted suggestions, the users can see the content matches in the training data by opening the Ansible panel within the editor.
This sends a \textit{contentmatches} request to the Ansible Lightspeed service which returns a list of content matches.
This list of content matches allows the user to review the sources which possibly match the suggestion thus improving the user confidence in the accepted suggestion.

\subsection{Inference Pipeline}

When the Ansible Lightspeed service receives a \textit{completion} request from the client, it has to return a code suggestion.
The completion request consists of a \textit{prompt} field and the value is the Ansible code before the cursor position where the inline suggestion is triggered in the code editor. 
Within the service, the \textit{completion} request processing is passed through three stages: 
\textit{pre-processing}, \textit{inference} and \textit{post-processing}. 

\paragraph{Pre-processing}

In the \textit{pre-processing} stage the prompt is checked for valid YAML syntax and the indentations are adjusted based on the requirements of the inference request.
After that, it is passed through Ansible Anonymizer~\cite{ansibleanonymizer} to remove any personal, identifiable information. 
The cleaned up prompt is then passed to the next stage where the inference request is sent to the model server. 

\paragraph{Inference}

IBM watsonx Code Assistant for Red Hat Ansible, or WCA-Ansible, is the model that powers Ansible Lightspeed.

WCA-Ansible is a transformer-based decoder trained from scratch on natural language, source code, and Ansible data.
We use the Hugging Face Transformers framework~\cite{huggingface} to train a 350 million parameters model with a context window of size 1024; this model utilizes fewer parameters than the model available to subscription holders. 
The model is pre-trained with about 143 billion tokens and then fine-tuned with about 500k natural language to Ansible task samples.
The training environment, hyper-paramters and fine-tuning data creation techniques are similar to Ansible Wisdom~\cite{pujar2023automated}.
The pre-training and fine-tuning data is license filtered to remove all data samples coming from files or repositories with restrictive licenses. 
We also filter the data to remove hate, abuse and profanity.
The pre-training data is also deduplicated at a file level.
The model server runs on machines containing A$100$ GPUs equipped with sufficient memory to 
restrict the end-to-end latency within $2$~seconds on average.
Every model suggestion is a multi-line Ansible task with 6.13 lines and 20.1 tokens on average.
\paragraph{Post-processing}

After the inference response is received by the Ansible Lightspeed service the response is again passed through the Ansible Anonymizer to cleanup any personal identifiable information that might be present in the model output.
To ensure the suggestion follows the good practices of Ansible content development defined by the Ansible community~\cite{redhatansiblecop}, the response is passed through Ansible Risk Insights or ARI tool~\cite{ari}. 
The ARI tool runs multiple custom rules to modify the suggestion as required to ensure best practices such as invoking modules using their fully-qualified collection names and reusing Ansible variables defined in the context.
It also redacts any credentials and removes and/or replaces any deprecated Ansible syntax from the suggestions. 
After post-processing is complete, the response is sent back to the client. 

\subsection{Content Matching Pipeline}

When the Ansible Lightspeed service receives a \textit{contentmatches} request from the client with the accepted suggestion as a payload, it triggers the following actions. 

The service encodes the recommendation using a Sentence-Transformers (SBERT~\cite{reimers2019sentence}) model.
The service sends the encoded suggestion text to a search instance that utilizes the k-nearest neighbor~(k-NN) algorithm~\cite{duda1973pattern} to return the top three nearest matches to the text suggestion. The search instance is indexed with the Ansible data used for fine-tuning the model.

Each match consists of information like the name and URL of the repository, the file path in the repository, the license, and the match score as per the k-NN algorithm.

\subsection{Data Collection and Analysis Framework}
\label{sec: analysis_framework}

\begin{figure}[t]
    \centering
    \includegraphics[width=\linewidth]{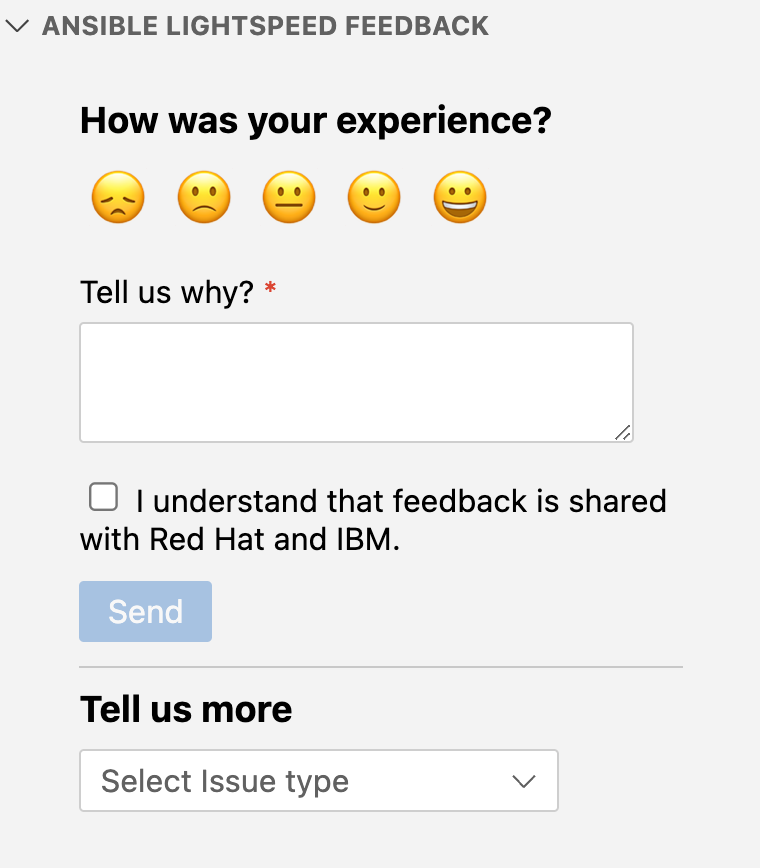}
    \caption{Ansible Lightspeed Feedback User Interface. The users can rate their experience using emojis, which we call star-rating. The right most emoji would indicate the best experience or a 5-star rating, and the left most emoji would indicate the worst experience or a 1-star rating. Users are also expected to write a few words to explain their rating in "Tell us why?" section.}
    \Description{}
    \label{fig:feedback_ui}
\end{figure}

Ansible Lightspeed collects explicit user feedback, and implicit event tracking.
For the explicit feedback, we have created a dedicated user interface (UI) in the sidebar of VS Code Editor as shown in Figure~\ref{fig:feedback_ui}. Within this UI, users can provide ratings on a scale of 1 to 5 stars, represented as emojis with rightmost emoji being 5 stars (the best) and leftmost emoji being 1 star (the worst).
Along with the rating the users also provide a written feedback on their experience using the service.
More on this explicit feedback in Sec.~\ref{sec: user-feedback}.

The event tracking platform monitors and records a wide range of events about user interactions, and system activities. This platform is a critical part of our data collection pipeline, ensuring that we systematically capture important events within our system. It includes user-initiated actions like completion requests, suggestion acceptance, as well as service-related events like server requests, matching sources, and post-processing logs.

The analysis framework consists of two parts: (1)~a usage analyzer evaluating engagement of the users, and (2)~a suggestion analyzer evaluating the quality of the code suggestions.

\paragraph{Usage analyser}

After collecting the telemetry data, we use event segmentation techniques to categorize and filter the data, allowing us to focus on specific event types and user behaviors. This helps us identify patterns, trends, and anomalies in the data, which in turn provides valuable insights into user preferences, satisfaction levels, and usage trends for Ansible Lightspeed.

\paragraph{Suggestion analyzer}
\label{subsubsec: suggestion-analyzer}

\begin{figure}[t]
    \centering
    \includegraphics[width=\linewidth]{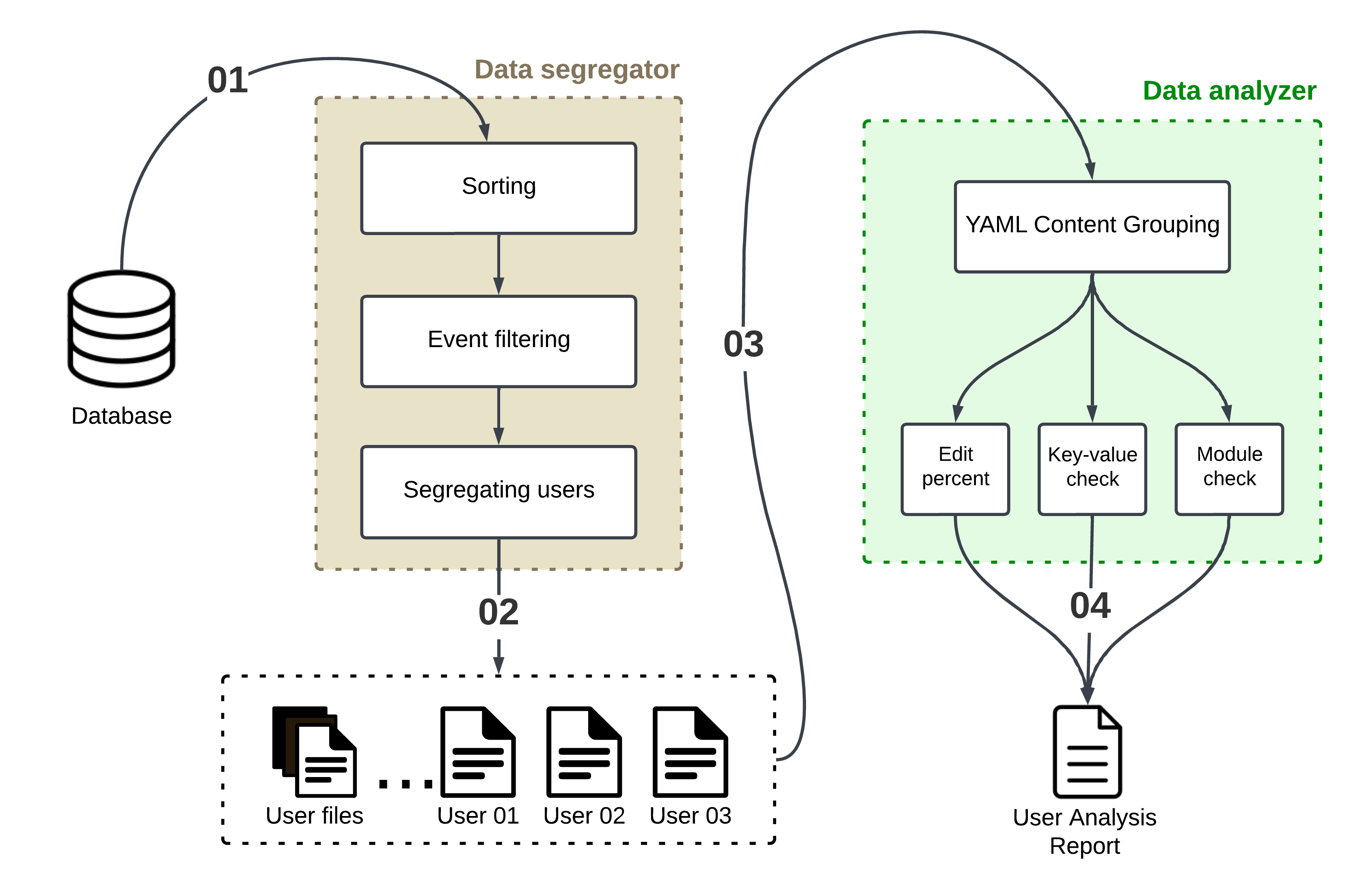}
    \caption{Architecture of the suggestion analyser.}
    \Description{}
    \label{fig:edit_analysis_framework}
\end{figure}

The second analysis is done on the user's actions following the acceptance of the suggestions. We designed a Python framework to gather the playbook content of each user to study the modifications made to the accepted task suggestions.
The architecture of the framework is shown in Figure~\ref{fig:edit_analysis_framework}.
%\todo{If we need to save space, remove Figure~\ref{fig:edit_analysis_framework}.}
It runs in two steps: data processing and data analysis.

We begin processing data by sorting user data and only keeping essential information related to three specific event categories: Suggestion events (regarding suggestions provided by the model), User action events (indicating whether the user accepted or rejected the suggestion), and Content events (related to document content). After organizing the data into separate files for each user, we send it to the analyzer for a detailed analysis of each user's data.

In the analyzer, we combine data from the sequential events explained above to create pairs of \textit{suggestion received} and \textit{suggestion committed}. Then, we break down the YAML content into lines and perform a sequence comparison. We use the SequenceMatcher class from the difflib library~\cite{difflib} to compute the deltas. SequenceMatcher uses gestalt pattern matching~\cite{gestalt}, which aims to identify the longest contiguous matching sub-sequence without “junk” elements. This recursive process is applied to both sides of the matching sub-sequence within the sequences. Although it does not provide minimal edit sequences, it often aligns well with human perceptual expectations in sequence comparisons. Finally, we parse the YAML into objects to gain insights into the parts where the user has made edits.

This analysis provides us with two crucial insights:
Firstly, it helps us determine if the user has entirely altered the suggestion, which is essential for assessing the actual acceptance of the suggestion.
Secondly, it allows us to identify which parts of the suggestion have been modified, giving us an idea of how well the suggestion aligns with the user's expectations.

\section{Analysis of User Interactions}
\label{sec: analysis}
The goal of our analysis is to determine if the users find the service useful and how we can improve the service to better fit user needs.
To achieve this, we try to address the following questions:
\begin{itemize}
    \item When do users use Ansible Lightspeed?
    \item Do users continue to use the service after trying it once?
    \item Do users modify accepted suggestions?
    \item At what rate do users accept suggestions?
    \item Why do users edit module names in accepted suggestions?
\end{itemize}

Our analysis is based on data collected from users who agreed to share their usage data of the closed beta and free Tech Preview version of the Ansible Lightspeed service from April 15, 2023, to May 15, 2024, a total of 397 days.

\newcommand{\tikzcircle}[2][red,fill=red]{\tikz[baseline=-0.5ex]\draw[#1,radius=#2] (0,0) circle ;}%

\definecolor{weekday}{RGB}{148,190,45}
\definecolor{weekend}{RGB}{215,1,74}

\begin{figure*}[t]
    \centering
    \includegraphics[width=0.9\linewidth]{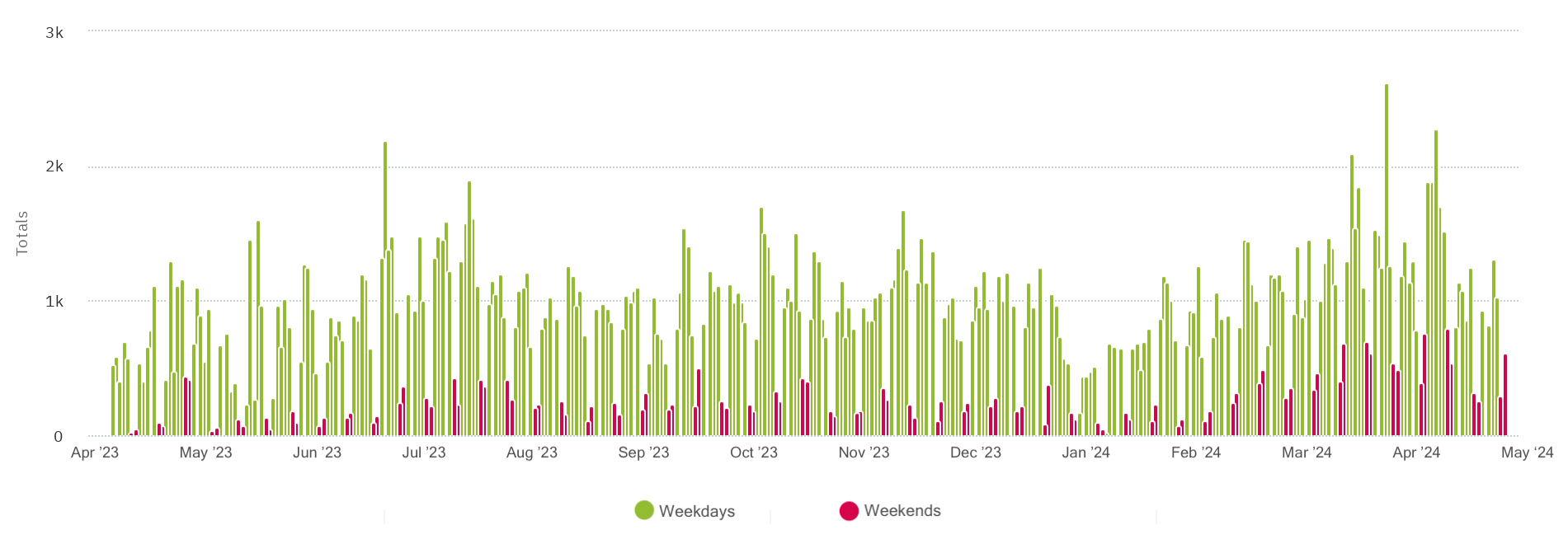}
    \caption{Number of Ansible Lightspeed completion requests per day. Weekdays are displayed in green [\tikzcircle[weekday, fill=weekday]{0.7ex}] and weekends in red [\tikzcircle[weekend, fill=weekend]{0.7ex}]. Average completion requests per day - Monday: 832.69, Tuesday: 1065.08, Wednesday: 990.92, Thursday: 1063.19, Friday: 864.77, Saturday: 241.25, Sunday: 264.90.
    The data is normalized based on different time zones from where the users made the completion requests.}
    \Description{}
    \label{fig:daily_completion_requests}
\end{figure*}

\subsection{Temporal Analysis}

Lightspeed's goal is to improve the productivity of developers who use Ansible for work.
Understanding when requests were made could give hint that Lightspeed is being used for work.

Figure~\ref{fig:daily_completion_requests} shows all Lightspeed completion requests\footnote{The figures here represent telemetry events, which may contain duplicates. The duplicates will inflate the request count but we believe the trends highlighted below will be remain the same.} on a daily basis for the entire analysis period.
Requests on weekdays are shown in green and those on weekends are shown in red.
One observation is that users tend to make more requests on an average weekday than on an average weekend.
We found that `Tuesdays' had the maximum average completion requests per day (average of 1065.08 completion requests) and `Saturdays' had the lowest average completion requests per day (average of 241.25 completion requests).
The reason for this weekday and weekend difference could be that a significant chunk of our users are using Lightspeed for work, during weekdays, rather than for study or side projects.
However, we need more analysis to be sure.

In Figure~\ref{fig:daily_completion_requests} we can also see a significant decline in the completion requests between December 23, 2023, and January 12, 2024. 
This could be due to the Holiday season when many offices are closed.

\begin{tcolorbox}
Q: When do users use Ansible Lightspeed?

\vspace*{0.5\baselineskip}

Users make more requests during weekdays, with noticeable drops on weekends and around holidays, which could mean that many users are using Lightspeed for work related requests.
\end{tcolorbox}

\subsection{User Retention}

\begin{figure}[t]
    \centering
    \includegraphics[width=\linewidth]{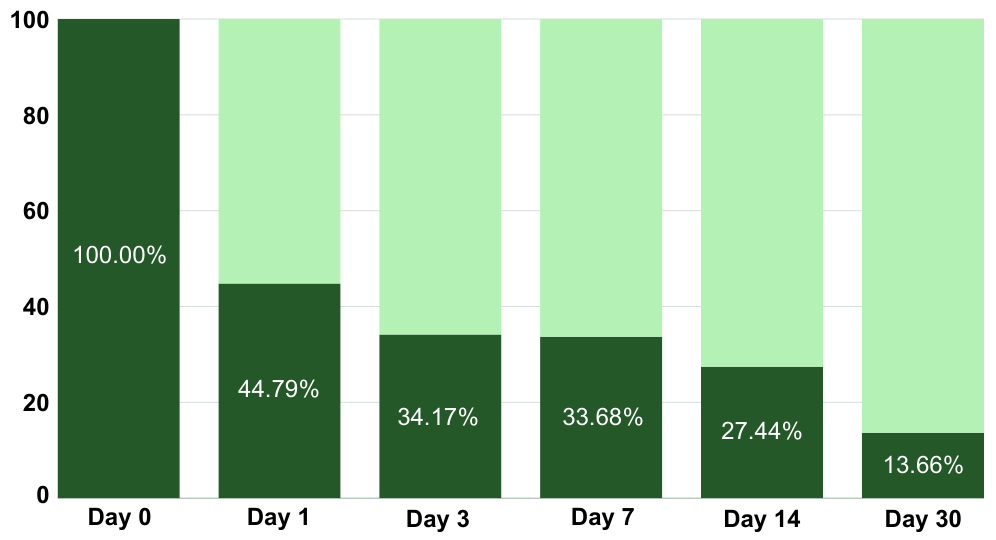}
    \caption{N-Day User Retention Trend of Ansible Lightspeed. The dark green represents the percentage of users returning on Day N.}
    \Description{}
    \label{fig:user_retention}
\end{figure}

\textit{User retention} metrics are crucial in-app and service analytics, providing valuable insights into user engagement and app longevity~\cite{user_retention}. 
Our study utilizes the N-day User Retention metric to evaluate the Ansible Lightspeed service~\cite{n_day_user_retention}.
This metric calculates the percentage of users who continue interacting with an app N days after their initial installation or first use.
Because of this metric's utility in serving as a barometer for user satisfaction, it is used in many industries, including by iOS and Android app developers~\cite{mobile_app_retention_2023}. 
For instance, iOS apps have an average Day 30 retention of 4.13\%, while the same number for Android apps is 2.59\%~\cite{app_retention_benchmarks_report_2022}.

Figure~\ref{fig:user_retention} shows the N-day user retention of Ansible Lightspeed service.
As can be seen, of all the users who use the service on their Day 0, 44.79\% return on Day 1, 34.17\% on Day 3 and so on. 
13.66\% users who first used the service 30 days prior, continue to use the service on Day 30.

Based on N-day user retention, we define a \emph{returning user}, as someone who returned to use the service at least once more after Day 0.
Out of a total of about 10,696 unique users, 3,910 or 36.6\% were returning users.
Since returning users are much more engaged with the service, we assume that they may be using the service for real work, rather than testing it using toy examples.
All further analysis is on this cohort of 3,910 are returning users.

\begin{tcolorbox}
Q: Do users continue to use the service after trying it once?
  \begin{itemize}
    \item On Day 1 the average retention rate is 44.79\%. On Day 30 it is 13.66\%.
    \item 3,910 users out of total 10,696 users, or about 36.5\%, are returning users.
  \end{itemize}
\end{tcolorbox}

\subsection{Edit Analysis}
\label{subsec: acceptance}

\begin{figure}
\begin{subfigure}{0.5\textwidth}
    \includegraphics[width=\textwidth]{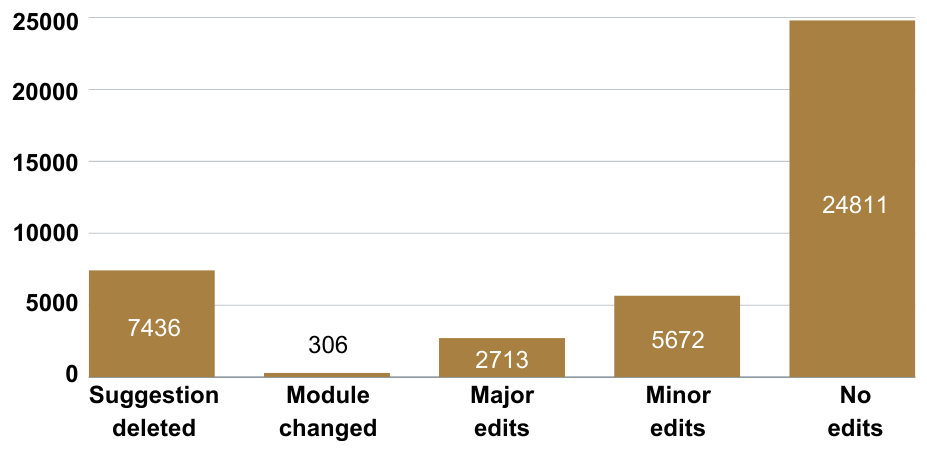}
    \caption{Distribution of suggestion modifications in the accepted suggestions. "Module Changed" indicates Minor edit suggestions in which module was changed.}
    \label{fig:edit_analysis_results_first}
\end{subfigure}
\vfill
\vspace{1em}
\begin{subfigure}{0.5\textwidth}
    \includegraphics[width=\textwidth]{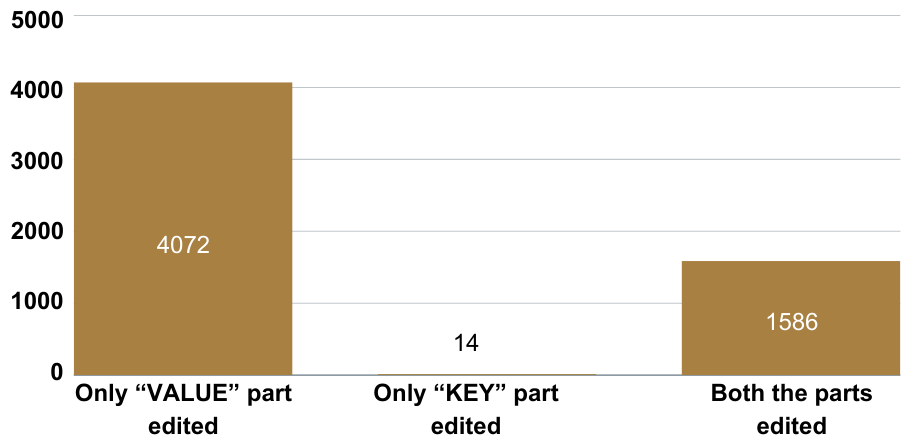}
    \caption{Distribution of user edit patterns in the scenarios where they have made minor edits in the accepted suggestions.}
    \label{fig:edit_analysis_results_second}
\end{subfigure}        
\caption{Ansible Lightspeed suggestion modification and edits analysis.}
\Description{}
\label{fig:edit_analysis_results}
\end{figure}

The interactions shown in Figure \ref{fig:ansible_lightspeed_flow} are recorded as telemetry events,\footnote{There are duplicate telemetry events. We can't be sure why there are duplicate events, but it may be as simple as network issues or user hitting the enter key twice, instead of once. We filter out duplicates and consider only unique requests for our analysis.} and this telemetry data is used to calculate the initial \textit{acceptance rate} of Lightspeed’s suggestions. 

We use the data of 3,910 returning users for our user acceptance rate analysis, since we consider returning users to be more engaged.
These returning users made a total of 62,099 completion requests to Ansible Lightspeed and received that many suggestions.
Out of 62,099 suggestions received by returning users, 40,938 suggestions were accepted, resulting in an initial acceptance rate of 65.92\%.

\begin{table}
  \caption{Summary of the user edit analysis. A returning user is a user who engaged with the system for more than 1 day. All numbers below, except Total users, are for returning users.}
  \label{tab:user_edit_analysis_numbers}
  \begin{tabular}{lr}
    \toprule
    Events& Numbers \\
        \midrule
    Total users&10,696\\
 Returning users &3,910\\
    Total suggestions& 62,099\\
    Average number of lines per suggestion & 6.13\\
    Average number of tokens per suggestion & 20.1\\
    Initially accepted suggestions& 40,938\\
   Fully accepted suggestions (0\% edits)& 24,811\\
 Minor edits (< 50\%) made after acceptance& 5,672\\
 Major edits ($\ge$ 50\%) made after acceptance& 2,713\\
  Deleted suggestions after acceptance& 7,436\\
 Module changed, but not major edit& 306\\
\bottomrule
  \end{tabular}
\end{table}
To understand if these accepted suggestions were used by users, we further examined if changes were made by users to the accepted suggestions.
Table~\ref{tab:user_edit_analysis_numbers} and Figure~\ref{fig:edit_analysis_results}.a gives a high level overview of the edits made by returning users on accepted suggestions.

In 24,811 instances~(60.60\% of the accepted suggestions), Ansible Lightspeed's suggestions were used without edits so we can assume that they aligned with user requirements. 
In 5,672 scenarios~(13.85\% of the accepted suggestions), users made small adjustments, less than 50\% edits,\footnote{Suggestion analyzer part of Sec.~\ref{subsubsec: suggestion-analyzer} explains how edits are calculated.} to the accepted suggestions.
An example of such a light or minor edit would be, when Ansible Lightspeed accurately predicts the correct module, users modified one or more module keys (options) and/or values to better align with their specific needs. 

In 2,713 instances~(6.62\% of the accepted suggestions), we observe substantial modifications (50\% edits or more) after acceptance, indicating that while users accepted the suggestion, they had to heavily modify it before they could actually use it.

We also noticed that users deleted 7,436 suggestions after acceptance, which is 18.16\% of the total initially accepted suggestions. 
Currently there is no way to tell why the users would delete accepted suggestions and move on to next request.
One explanation is that users found it challenging to review the suggestion in the greyed text (inline suggestion text) within the VS Code editor, leading them to initially accept and then delete after closer review.
Another reason could be that they were experimenting with different prompts trying one after another and discarding the answers.

We further analyzed 5,672 instances which were lightly edited, to make sure that they were indeed minor edits.
Figure~\ref{fig:edit_analysis_results}.b gives a break down of different kinds of minor edits.

The first observation was that there were 306 instances of minor edits, or 5.4\% of all accepted suggestions, in which the module name was also edited.
A module could not be considered a minor edit because the keys and values which immediately follow depend on the type of module.
If the model gets the module wrong, it is more likely to get rest of the suggestion wrong as well.
If the module name is edited, we consider the suggestion to be rejected.

We found that in a majority of remaining instances, the users opted to change only the `VALUE' component of the received suggestions. This involves personalized modifications to variable names, numeric values, and strings, something which maybe hard for Lightspeed to predict based only on the name field prompt and preceding context.

In a much smaller proportion users made changes in both the `KEY' and the `VALUE' parts of the suggested options of the module, added a pair option and value, or removed one.
This kind of edit slightly changes the behavior of the task while staying within the context of the suggested module. 
Instances in which only the 'KEY' was changed were negligible.

\begin{tcolorbox}
Q: Do users modify accepted suggestions?
  \begin{itemize}
    \item Yes users modify accepted suggestions.
    \item 18.16\% of initially accepted suggestions were deleted.
    \item 6.62\% initially accepted suggestions were heavily modified.
  \end{itemize}
\end{tcolorbox}

\subsection{User Acceptance Rate}
The edit analysis described above shows that the acceptance rate as currently defined can be misleading.
Which is why we define a new acceptance rate metric, called \emph{Strong Acceptance rate}. 
Strong Acceptance rate is the percentage of accepted suggestions, where a suggestion is considered accepted only if:
\begin{enumerate}
\item The suggestion is initially accepted by the user.
\item After acceptance, the suggestion is not deleted.
\item Suggestion has minimal edits or less than 50\% of the suggestion is edited.
\item Critical components, like the module name in case of Ansible, should not be edited.
\end{enumerate}

As per this definition, the Strong Acceptance rate for Ansible Lightspeed comes out to 49.08\%, compared to the initial acceptance rate of~65.92\%.
We feel Strong Acceptance rate more closely aligns with users perception of what they consider accepted, compared to the initial acceptance rate.

Table.~\ref{tab:acceptance_rate_comparison} shows Ansible Lightspeed acceptance rate compared to other IDE based code completion tools.
We compared our acceptance rate with the initial acceptance rates of Google Machine Learning enabled code completion system~\cite{googledevprod} (MLECC) which is an internal tool at Google, Github CoPilot~\cite{ziegler2022productivity} and CodeCompose~\cite{murali2023codecompose}, which is an internal tool at Meta. 
These were the only other code completion tools we found to have used acceptance rate as a metric.
%Tab.~\ref{tab:acceptance_rate_comparison} presents the comparison.
Google MLECC~\cite{googledevprod} provides acceptance rates for single-line and multi-line code completion, but don't breakdown acceptance rate based on languages. 
\citet{ziegler2022productivity} for GitHub Co-pilot and \citet{murali2023codecompose} for CodeCompose share acceptance rate for individual languages.
We show the top three languages with best acceptance rate.
The acceptance rate for these three baselines is what what we call the initial acceptance rate. 
It merely tells us whether the user accepted the suggestion offered without telling us how the user actually used the suggestion.
It does not give any insight into whether the user accepted the suggestion as is, or edited the suggestion or altogether deleted it. 

Ansible Lightspeed always generates multi-line Ansible code suggestions, with an average of 6.13 lines and 20.1 tokens per suggestion over all the suggestions shown in Table.~\ref{tab:user_edit_analysis_numbers}.
Lightspeed has a higher acceptance rate compared to the other tools for both initial accept and Strong Acceptance rate.

\begin{table}
  \caption{Comparison of acceptance rate of Ansible Lightspeed with the other code completion tools. Google MLECC~\cite{googledevprod} stands for ML enabled code completion. Ansible Lightspeed Initial accept are percent suggestions which were accepted by users. Strong Acceptance rate is percent suggestions which were accepted and had less than 50\% user edits, based on difflib SequenceMatcher\cite{difflib}.}
  \label{tab:user_acceptance_rate_comparison}
  \begin{tabular}{lcc}
    \toprule
    Service & Acceptance Rates & User Count\\
    \midrule
    \textbf{Google MLECC} &  & \\
    single-line CC & 25\% & 10k+\\
    multi-line CC & 34\% & 5k+\\
    \midrule
    \textbf{Github Co-Pilot} &  & 17,420\\
    Typescript & 26.3\%  & \\
    Javascript & 31.4\% & \\
    Python & 30.8\% & \\
    Other & 23.4\% & \\
    \midrule
    \textbf{CodeCompose} & & \\
    Hack & 22.5\% & 5.5k\\
    Python & 22.0\% & 10.7k\\
    C & 21.3\% & 201\\
    All & 22\% & 16k\\
    \midrule
    \textbf{Ansible Lightspeed} &  & \\
    Initial accept (Ansible) & \textbf{65.9\%} & 3,910\\
    Strong accept (Ansible) & \textbf{49.1\%} & 3,910\\
    \bottomrule
  \end{tabular}
  \label{tab:acceptance_rate_comparison}
\end{table}

\begin{tcolorbox}
Q: At what rate do users accept suggestions?
  \begin{itemize}
   \item Initial acceptance rate is misleading since users can delete or heavily modify suggestions.
    \item 65.9\% of suggestions are initially accepted.
    \item We define new acceptance rate, Strong Acceptance rate.
    \item 49.1\% of model suggestions were accepted as per Strong Acceptance rate.
    \item Ansible Lightspeed has relatively higher acceptance for Ansible, compared to more general tools for various programming languages.
  \end{itemize}
\end{tcolorbox}

\subsection{Edited Module Analysis}

There are a total of 1,710 initially accepted suggestions which were explicitly edited by the users at the module level.
We consider such suggestions to be rejected for Strong Acceptance rate.
However, the users must have found something useful in the suggestion for them to accept it and then spend sometime editing it.

We analyse these edits in order to better understand how the model suggestions differ from user expectations.
The edit categories were identified based on manual analysis. 
The number of samples in each category was obtained based on rules to find similar samples.
A suggestion could have been edited for multiple reasons, so an edited sample can belong to multiple categories.

\subsubsection*{User preference}
In order to remove ambiguity in module names, Ansible Lint encourages users to use Fully Qualified Collection Name, or FQCN~\cite{ansible_fqcn}, when declaring modules.
In keeping with this best practice, Ansible Lightspeed always generates module names in FQCN format.
On the other hand it is quite possible that some users prefer the shorter version of the module name.
For example, the best practice FQCN reference would be \emph{ansible.builtin.debug} while some users may prefer to refer to the same module as \emph{debug}.

We find that in 346, or 20.2\% of, instances the users edited the model suggestion and changed the FQCN module name to it's shortened format. 
The body of the generated Ansible task often had little to no edits, indicating that the FQCN module name, which is correct, is not what they prefer.

\subsubsection*{YAML reorganization}
As per our observation, in the process of writing an Ansible playbook users often reorganize the YAML structure.
This reorganization involves adding new valid Ansible keys like \emph{block}, \emph{tag}, \emph{register}, \emph{loop} and \emph{become}.
The model generated suggestion is generally part of the reorganized playbook, either in the edited or original format, and is likely to have inspired the reorganization.

Such edits indicate that the generated suggestion is not necessarily incorrect, but is definitely incomplete and can be better organized. 
There are at least 211 (12.3\%) such instances of reorganization.

\subsubsection*{Functionally similar modules}
% multiple functionally similar modules
Module confusion usually occurs when there are multiple functionally similar modules which can be used to implement user's intent and the user is at least somewhat familiar with one of them.

One example of this is \emph{command} and \emph{shell} modules, which are often interchangeable.  
In addition, functionalities of many other modules can be implemented by \emph{command} or \emph{shell} modules.
It is possible that users are not aware of other modules, or they may prefer another module than the one that lightspeed generated.
Generated shell commands are also likely to be incorrect, which could inspire more user edits.
%Another observation is that the model often generates incorrect shell command which may also cause the user the reject or heavily edit the suggestion.
There are 352 edits (20.6\%) which are related to the command/shell module.
Apart from \emph{command} or \emph{shell} module related edits, we found 336 (19.6\%) instances of edits related to modules which have similar functionalities and can be used interchangeably. 
%Full list and examples can be found in Appendix. \ref{sec: user_edit}.

There are another 586 (34.2\%) cases where the user edited the module. 
Understanding these edits requires more analysis.

\begin{tcolorbox}
  Q: Why do users edit module names in accepted suggestions?
  \begin{itemize}
    \item Suggestion is edited to remove FQCN based on user preference.
    \item YAML is reorganized to add more details which the model suggestion is missing.
    \item Certain functionalities can be implemented in multiple ways. Users change the suggestion based on their preferred implementation.
  \end{itemize}
\end{tcolorbox}

\section{User Feedback}
\label{sec: user-feedback}

A total of 605 users have provided rating and comments using the the sidebar for Ansible Lightspeed in the VS Code Editor~(Figure~\ref{fig:feedback_ui}).

\subsection{Star rating}
\begin{figure}
    \includegraphics[width=\linewidth]{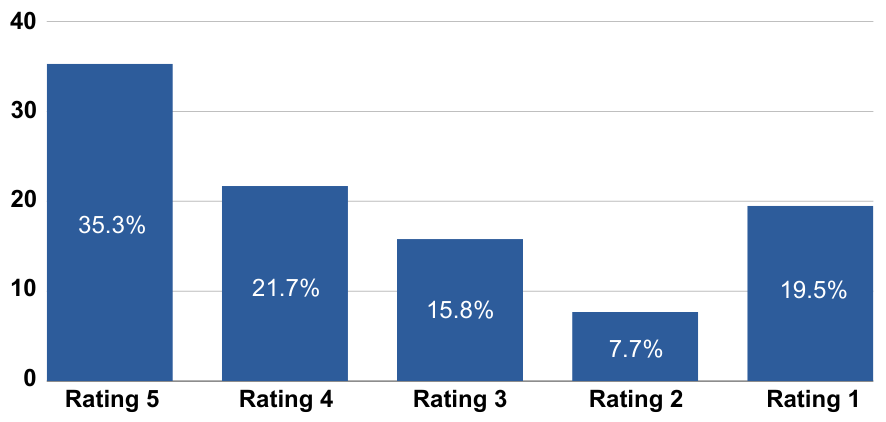}
    \caption{Ansible Lightspeed's user sentiments.}
    \Description{}
    \label{fig:relative_sentiment}
\end{figure}

A breakdown of ratings in terms of stars is presented in Figure~\ref{fig:relative_sentiment}.
Notably, a total of 57\% of users express considerable satisfaction with the service, as indicated by their assignment of 4 and 5-star ratings.
The 3-star rating by 15.8\% of users indicates a moderate level of satisfaction.
In this case users do find some value in using Ansible Lightspeed, but expect more improvement.
About 27.2\% users appeared dissatisfied and gave a rating of 2 and 1-star.

\subsection{User Comments}
Apart from star-rating, users also submit feedback in the form of textual comments.
We manually review and classify negative feedback (1 or 2 stars) and positive feedback (4 or 5 stars). 
The categories are discovered in the data using open coding.
Each comment is assigned to only one category, the main point of the feedback.
We also share some examples of real user comments, both positive and negative.
These quotes have been rephrased for clarity and anonymity, removing any references to user details.

\paragraph{Negative feedback}

\begin{figure}
    \includegraphics[width=\linewidth]{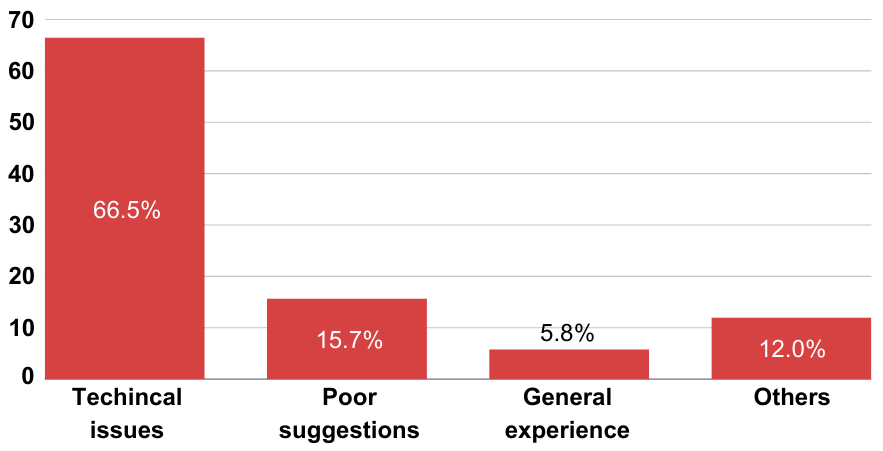}
    \caption{Distribution of negative feedback.}
    \Description{}
    \label{fig:negative_feedback_distribution}
\end{figure}

The breakdown of negative comments based on manual review is presented in Figure~\ref{fig:negative_feedback_distribution}.
The majority of the negative feedback, 66.49\%, is coming from users that do not manage to get Ansible Lightspeed to work. These problems stem from various sources, including installation difficulties, authentication troubles, network problems, and bugs.

The second largest source of complaints, accounting for 15.71\%, is due to poor suggestions provided by the model. This could sometimes be caused by the use of uncommon modules, as illustrated by the following example of feedback:
\begin{quote}
    \textit{“I mainly use a specific collection that may not be part of training so suggestion were rarely helpful”}
\end{quote}

Some users are disappointed with the suggestions provided by the model because their requests fall outside of Lightspeed's current scope.
For instance, they may ask for multi-task or playbook generation, but the model is currently only trained for generating single tasks:
\begin{quote}
    \textit{“Has really bad recommendations. Tried creating a playbook with name "Get ..." and it get me a module recommendation”}
\end{quote}

Finally, 5.79\% of users are dissatisfied with their overall experience, specifically citing the need to re-authenticate periodically as an issue. 

The remaining 12.04\% of comments are not informative.

\paragraph{Positive feedback}

\begin{figure}
    \includegraphics[width=\linewidth]{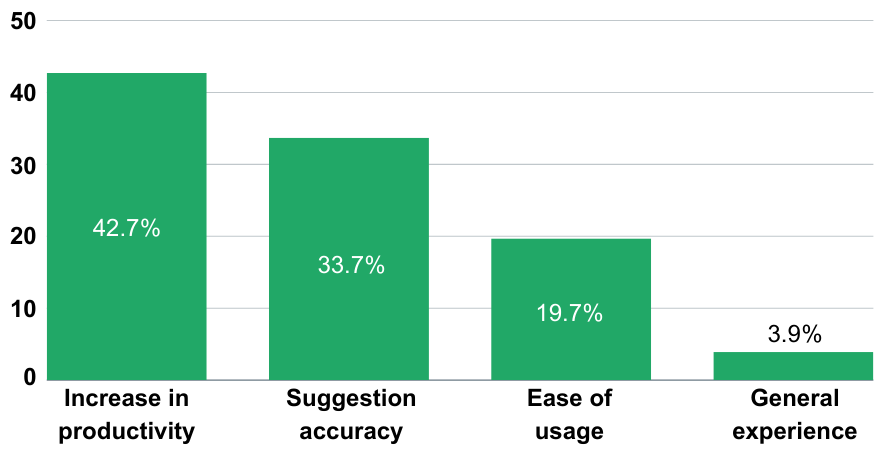}
    \caption{Distribution of positive feedback.}
    \Description{}
    \label{fig:positive_feedback_distribution}
\end{figure}

Breakdown of reasons for positive experience is presented in Figure~\ref{fig:positive_feedback_distribution}.
About 42.7\% of users consider improved productivity as Ansible Lightspeed's key benefit.
Multiple users mentioned that they noticeably saved time by using Lightspeed.
\begin{quote}
   \textit{ “Lightspeed is genuinely amazing and has significantly impacted my productivity. It saves me from the hassle of constantly switching between Ansible documentation websites for reference.”}
\end{quote}

\begin{quote}
    \textit{“This extension is quite handy for speeding up my daily work in the office. It provides helpful task suggestions, and the inline module suggestions really cut down on time spent.”}
\end{quote}
33.7\% of users found suggestion accuracy to be the most impressive part of Ansible Lightspeed. Users have commented that the model not only recommends the correct module but also the associated module options and values.
\begin{quote}
    \textit{“This was beyond my expectation that the tool could manage complex syntax without any issues. Also, the NLP capabilities are impressive. I tested it with different writing styles for the same prompts, and it consistently provided the expected results. For example, it handled both of these cases perfectly: \newline
    $1$ - name: Can we see the value of win\_credential\_results?
    $2$ - name: Print win\_credential\_results”}
\end{quote}

About 19.7\% of customers praise the service's user-friendliness, highlighting its simplicity in setup and use. 
For instance, some users have mentioned the login and on-boarding process as being easy and fast.
Remaining 3.9\% of the users said best part was the general experience of using the tool.

\section{Related Work}
\label{sec: related}
LLMs have shown remarkable ability to generate code, with many recent models like GPT-4~\cite{achiam2023gpt}, Llama~\cite{touvron2023llama}, StarCoder~\cite{li2023starcoder} and others performing very well on code evaluation benchmarks like HumanEval~\cite{chen2021evaluating} and MBPP~\cite{austin2021program}.

A lot of work has also been done on developing coding assistants for IDEs based on LLMs.
These would include tools that are internal to a company and are only accessible by internal users and those that are openly available.
\citet{murali2023codecompose} analyze CodeCompose, an internal code completion tool at Meta which is based on InCoder LLM \cite{fried2022incoder}, and show that the acceptance rate is about 22\% across 9 programming languages for approximately 16k users.
Similarly, \citet{googledevprod} analyze a code completion tool at Google and measure an acceptance rate of 25-34\% over 10k+ Google-internal developers.
\citet{svyatkovskiy2020intellicode} introduce and evaluate IntelliCode Compose on multiple programming languages, but they use evaluation based on edit-distance compared to the ground truth and do not mention any user acceptance metrics or the total number of users.
None of these works mention user sentiment or user retention over an extended period of time.

The openly available coding assistants would be GitHub Copilot~\cite{GitHubCopilot}, Tab9~\cite{Tab9}, Replit~\cite{replit} and Amazon CodeWhisperer~\cite{codewhisperer} among others. 
Among these, GitHub Copilot has been widely studied since its release.
\citet{nguyen2022empirical} test Copilot on 33 LeetCode questions in four programming languages. 
\citet{vaithilingam2022expectation} perform a more user-centred evaluation with 24 users to see how programmers use and perceive Copilot.
\citet{ziegler2022productivity} perform an in-depth study of user acceptance, similar to what we do, but for multiple programming languages and show that the acceptance rate of Copilot suggestions for different user categories, for different programming languages, is approximately 20\%-30\%.
\citet{peng2023impact} study the impact of Co-pilot on the speed of programmers and find that AI pair programmers are 55.8\% faster in implementing an HTTP server in JavaScript. 
However, they do not provide information about the acceptance rate of AI pair programmers.
\citet{yeticstiren2023evaluating} perform a comparative study of GitHub Copilot, Amazon CodeWhisperer and ChatGPT \cite{chatgpt} (sibling model of InstructGPT~\cite{ouyang2022training}) in terms of code quality metrics like code correctness, code security, code reliability and code maintainability but do not consider acceptance of code by real users.

We did not find any user retention figures for any of the existing code completion systems or for the VS Code plugins in the VS Code marketplace.

\section{Limitations}
\label{sec: limitations}

While this analysis provides comprehensive insights into the impact of Ansible Lightspeed on user experience, there are certain considerations and limitations to be acknowledged.

We do not perform acceptance rate and edit analysis on all Ansible Lightspeed users.
Our Telemetry data includes events from users who may not be Ansible programmers and simply want to try out the system with toy examples, which could add noise to our analysis. 
To mitigate this issue, we created a cohort of returning users who used Ansible Lightspeed on two or more days.

Ansible Lightspeed's acceptance rate is compared with models which are much more general, can be used for multiple programming languages and are likely bigger in terms of number of parameters and the amount of data used for training.
These general models also tend to have a higher number of users overall and their performance on Ansible-YAML has not been studied or reported.
One of the main purposes of our study was to explore the performance of a domain, language-specific model vis-a-vis a more general model with real users.
The total number of users, 3,910, is sizable enough for us to be confident about our analysis.

Additionally, our evaluation of Ansible Lightspeed's user retention lacked official data for comparisons with other code generation services. 
Lacking user retention baseline for code completion tools, we can't say with confidence that users find Ansible Lightspeed useful enough to keep using it, especially for work.
This is why we rely on temporal trends, acceptance rates, and explicit feedback to make such assessments.

Despite these considerations, we hope the analysis and insights presented here offer a substantial understanding of Ansible Lightspeed's effectiveness, user engagement, and potential areas for improvement. Based on this work, we hope to continue investigations into domain-specific code generation tools and their impact on IT automation.

\section{Conclusion and Future work}
\label{sec: conclusion}
\citet{pujar2023automated} show that a relatively small model fine-tuned on a specific Ansible task generation use case can outperform a much larger model on a benchmark dataset.
In this paper, by analyzing user interaction data of 3,910 users, we show that the suggestions made by such a specifically fine-tuned model can have a relatively higher acceptance rate among a large set of users.

Analysis of feedback from 605 users and of 1,710 user edits to Lightspeed generated Ansible tasks led to interesting findings which we plan to use to improve Ansible Lightspeed.
Currently, we use greedy decoding to generate one suggestion for every prompt.
To account for user preferences and the fact that there can be multiple modules that can be used to implement similar functionalities, we hope to generate multiple suggestions.
Also, currently we support only Ansible task generation, but we are working to expand the capabilities of the model to be able to generate longer Ansible sequences.
With this, we hope to reduce the need for the user to reorganize the code and add missing details.
We also plan to develop capabilities for Ansible code explanation, debugging, and as well as customization for specific users.

We hope that our proposed user edit-based acceptance criteria, Strong Acceptance rate, is used by other code completion tools to get a better estimate of whether users truly found the model suggestion helpful.
We also hope that by sharing Ansible Lightspeed's user retention figures, others get encouraged to share similar figures for their code completion tools.
And finally, by sharing the code for Ansible lightspeed service\footnote{https://github.com/ansible/ansible-ai-connect-service} and our analysis framework\footnote{https://github.com/ansible-community/lightspeed-analysis-framework} we hope to help others trying to develop similar tools.

\begin{acks}
We want to thank the following people for their help, support, and consultation while envisioning, building and deploying Ansible Lightspeed at Red Hat:  Mattew Jones, Xavier Lecauchois, Brian King, Robin Bobbit, Scott Harwell, Craig Brandt, James Wong, Marty Turner, Goneri Le Bouder, Tami Takamiya, Sumit Jaiswal.

We would also like to thank the model development team and the Ansible Risks Insights team at IBM Research as well as the Watson Code Assistant team.

\end{acks}

\bibliographystyle{ACM-Reference-Format}
\bibliography{custom}

\end{document}